\documentclass[useAMS,usenatbib]{mn2e}
\usepackage{graphicx}
\usepackage{mathrsfs,amssymb}
\usepackage{amsmath}

\newcommand       \cm           {\,{\rm cm}}

\newcommand       \erg          {\,{\rm erg}}
\newcommand       \eV           {\,{\rm eV}}
\newcommand       \g            {\,{\rm g}}
\newcommand       \K            {\,{\rm K}}

\newcommand       \s            {\,{\rm s}}

\newcommand       \um           {\mu{\rm m}}
\newcommand       \mum          {\,{\rm \mu m}}
\newcommand       \Teff         {T_{\rm eff}}

\newcommand       \simali       {\sim\,}

\newcommand       \kabs         {\kappa_{\rm abs}}
\newcommand       \kabsint      {\kappa_{\rm abs}^{\rm int}}

\newcommand       \md           {M_{\rm d}}
\newcommand       \Td           {T_{\rm d}}
\newcommand       \xstar        {{\small \left[{\rm X/H}\right]}}
\newcommand       \Timin        {{\small \left[{\rm Ti/H}\right]_{\rm min}}}

\newcommand       \Tistar       {{\small \left[{\rm Ti/H}\right]_\star}}
\newcommand       \Smin        {{\small \left[{\rm S/H}\right]_{\rm min}}}
\newcommand       \Sstar       {{\small \left[{\rm S/H}\right]_\star}}
\newcommand       \Sistar       {{\small \left[{\rm Si/H}\right]_\star}}
\newcommand       \SitoH        {{\small \left[{\rm Si/H}\right]}}
\newcommand       \Femin        {{\small \left[{\rm Fe/H}\right]_{\rm min}}}
\newcommand       \Festar       {{\small \left[{\rm Fe/H}\right]_\star}}
\newcommand       \Cstar       {{\small \left[{\rm C/H}\right]_\star}}
\newcommand       \Castar       {{\small \left[{\rm Ca/H}\right]_\star}}
\newcommand       \CtoH        {{\small \left[{\rm C/H}\right]}}
\newcommand       \Ostar       {{\small \left[{\rm O/H}\right]_\star}}
\newcommand       \OtoH        {{\small \left[{\rm O/H}\right]}}
\newcommand       \XtoH        {{\small \left[{\rm X/H}\right]}}
\newcommand       \ppm          {\,{\rm ppm}}
\newcommand       \nx           {n_{\rm\small X}}
\newcommand       \mH           {m_{\rm\small H}}
\newcommand       \mHstar       {M_{\rm\small H}}
\newcommand       \mud          {\mu_{\rm d}}
\newcommand       \mdmax        {\left(\md/M_{\rm H}\right)_{\rm max}}
\newcommand       \mdmin        {\left(\md/M_{\rm H}\right)_{\rm min}}
\newcommand       \Etot         {E_{\rm tot}}
\newcommand       \lambdap      {\lambda_{\rm p}}

\title[On the Carriers of the 21 Micron Feature]
{On the Carriers of the 21 Micron Emission Feature in
Post-Asymptotic Giant Branch Stars}

\author[Zhang, Jiang \& Li]
{Ke Zhang$^{1}$, B.W. Jiang$^{1,2}$,
        and Aigen Li$^{2}$\thanks{%
                E-mail: zhangke@mail.bnu.edu.cn, bjiang@bnu.edu.cn,
                        lia@missouri.edu}\\
$^{1}$Department of Astronomy, Beijing Normal University,
                  Beijing 100875, China\\
 $^{2}$Department of Physics and Astronomy,
             University of Missouri,
             Columbia, MO 65211, USA}

\begin{document}

\maketitle

\begin{abstract}
The mysterious 21$\mum$ emission feature seen in sixteen C-rich
proto-planetary nebulae (PPNe) remains unidentified since its
discovery in 1989. Over a dozen of materials are suggested as the
carrier candidates. In this work we quantitatively investigate eight
inorganic and one organic carrier candidates in terms of elemental
abundance constraints, while previous studies mostly focus on their
spectral profiles (which could be largely affected by grain size,
shape and clustering effects). It is found that: (1)  five
candidates (TiC nanoclusters,
    fullerenes coordinated with Ti atoms, SiS$_2$,
     doped-SiC, and SiO$_2$-coated SiC dust)
    violate the abundance constraints
    (i.e. they require too much Ti, S or Si to account
    for the emission power of the 21$\mum$ band,
(2) three candidates (carbon and silicon mixtures,
    Fe$_2$O$_3$, and Fe$_3$O$_4$),
    while satisfying the abundance constraints,
    exhibit secondary features which are not detected in
    the 21$\mum$ sources, and
(3) nano FeO, neither exceeding
    the abundance budget nor producing undetected secondary features,
    seems to be a viable candidate,
    supporting the suggestions of \citet{Posch2004ApJ...616}.

\end{abstract}
\begin{keywords}
(stars:) circumstellar matter -- infrared: stars -- stars: AGB and
post-AGB -- stars: individual (HD\,56126)
\end{keywords}

\section{Introduction\label{sec:intro}}
Since its first detection in 1989
\citep{Kwok...1989ApJ...345L..51K}, the so-called ``21$\mum$ feature''
has been identified in sixteen proto-planetary nebulae (PPNe;
\citealt{Kwok1999IAUS..191}, \citealt{Hrivnak2009arXiv0902.0077H})
[and arguably also in two planetary nebulae (PNe) associated with
Wolf-Rayet central stars \citep{Hony2001A&A...378L} and in two
highly evolved carbon stars \citep{Volk2000ApJ...530}]. This feature
has little shape variation among different sources with a peak
wavelength at $\simali$20.1$\mum$ and a FWHM of
$\simali$2.2--2.3$\mum$. Most of these sources exhibit quite uniform
characteristics: they are metal-poor, carbon-rich F and G
supergiants with strong infrared (IR) excess and over abundant
s-process elements \citep{Zhang...2006PABei..24...43Z}.

After its discovery, over a dozen of carrier candidates have been
proposed (see Fig.~\ref{fig:kabs_21um}), with the number of proposed
candidate materials comparable to the total number of the 21$\mum$
feature sources (see \citealt{Andersen2005ESASP.577..447},
\citealt{Zhang...2006PABei..24...43Z}). These include both inorganic
materials:
    (a) TiC nanoclusters \citep{vonHelden...2000Sci...288..313V},
    (b) SiS$_2$ grains \citep{Goebel1993A&A...278},
    (c) Doped-SiC dust\citep{Speck2004ApJ...600},
    (d) carbon and silicon mixtures \citep{Kimura2005crystal...275} ,
    (e) SiC core-SiO$_2$ mantle grains \citep{Posch2004ApJ...616},
    (f) FeO \citep{Posch2004ApJ...616},
    (g) Fe$_2$O$_3$, and
    (h) Fe$_3$O$_4$ \citep{Cox1990A&A...236L};
and organic materials:
   (i) large-cage carbon particles (fullerenes)
       coordinated with Ti atoms
       \citep{Kimura2005ApJ...632},
   (j)   urea or thiourea \citep{Sourisseau1992A&A...254L},
   (k)   polycylic aromatic hydrocarbon (PAH), and
   (h)  hydrogenated amorphous carbon (HAC)
        \citep{Buss1993ApJ...415,Justtanont1996A&A...309}.

However, the carrier of this feature remains unidentified (see
\citealt{Posch2004ApJ...616} for an excellent overview). It is
considered as one of the most interesting unresolved mysteries in
astrochemistry \citep{Kwok2002ApJ...573}.
Previous studies mostly rely on a comparison of the spectral profile
of a candidate material with the observed 21$\mum$ emission feature.
In this respect, astronomers often consider spherical dust using Mie
theory. However, the spectral profile is expected to vary with grain
shape, size, and the presence of voids (e.g. see
\citealt{Huffman1989IAUS..135..329H,Li2008MNRAS.391L..49L,Voshchinnikov2008A&A...483L...9V}).
Although models based on spherical dust may not fit the observed
21$\mum$ emission feature, one can not rule out the possibility that
a reasonably good fit may be obtained by fine-tuning dust shape and
size distributions. Indeed, some studies have demonstrated that the
spectral fit is improved with a continuous distribution of
ellipsoids (CDE; see e.g. \citealt{Posch2004ApJ...616}) for the dust
shape. We should note that even the failure of CDE models can not
readily rule out the considered candidate material since the
spectral profile could be further affected by many other factors
(e.g., clustering of individual grains, see
\citealt{Rouleau1991JRASC..85..201R}).

In this work we attempt to constrain the nature of the 21$\mum$
feature carrier by performing a systematic study of the validity of
the above-listed candidate carriers. In view of the fact that a
spectral profile ``mismatch'' could probably be improved by
considering dust shape and size distributions, clustering of
individual grains, and the presence of voids, unlike previous
studies, we do not completely rely on whether the interested
candidate material produces a band profile closely matching the
observed one. Instead, we take an alternative approach which at
least complements previous approaches (which are mainly based on
spectral profile matching): we focus on the elemental abundance
required to account for the total power emitted in the 21$\mum$
feature. This approach is less affected by the (unknown) dust
clustering and shape distribution and is therefore more robust.

This paper is organized as follows: we first present in
\S\ref{sec:constraints} the general constraints on the validity of a
proposed carrier. We then apply these constraints to the
above-listed candidate materials (\S\ref{sec:candidate}). For this
purpose, we choose the prototypical 21$\mum$ feature source
HD\,56126 as a comparison basis. The main results are summarized in
\S\ref{sec:summary}.

\section{General Constraints: Band Strength
         and Abundance Budget}\label{sec:constraints}
For a candidate carrier to be a viable explanation of the 21$\mum$
feature, in addition to the close match with the 21$\mum$ feature
observed in PPNe, it must satisfy the abundance constraint,
i.e., the candidate material must be abundant enough
to account for the total power emitted in the 21$\mum$ feature.
Further, it should of course not produce any secondary features
which are not observed in the 21$\mum$ sources.

The 21$\mum$ feature is one of the strongest IR dust features in
C-rich evolved objects. The strongest 21$\mum$ source (HD\,56126)
emits $\simali$8\% of its total IR power in this feature, while the
21$\mum$ feature in other sources amounts to $\simali$1--7\% of
their total IR power \citep{Hrivnak2000ApJ...535..275H}. That the
21$\mum$ feature emits such a large energy would put stringent
constraint on the abundance of the carrier, particularly, on the
abundance of the relatively rare element in the suggested carrier
material (e.g. Ti in TiC).

Let $\Etot$ be the total power emitted in the 21$\mum$ band, which
is an observational parameter. It is related to the total mass of
the carrier dust $\md$ through
\begin{equation}
\Etot = \md \int_{21\mum\,{\rm band}} \kabs(\lambda) \times 4\pi
B_{\lambda}(\Td)d\lambda ~,
\label{equation:etot}
\end{equation}
where the integration is over the 21$\mum$ band (but with the
continuum underneath the 21$\mum$ feature subtracted), $\kabs$ is
the mass absorption coefficient (also known as ``opacity'') of the
dust, $B_\lambda(T_{\rm d})$ is the Planck function of a black body
with temperature $T_{\rm d}$ at wavelength $\lambda$, and $\Td$ is
the dust temperature.

If the carrier material contains element X,  the abundance of
element X (relative to H) in the dust can be written as
\begin{equation}
\xstar = \frac{\nx\,\md/\mud \mH}{\mHstar/\mH} ~,
\label{eq:XtoH}
\end{equation}
where $\mHstar$ is the total hydrogen mass of
the circumstellar envelope of the object,
$\mH$ is the mass of a hydrogen atom, $\mud$ is the
molecular weight of the dust grain, and $\nx$ is the number of atoms
per molecule for element X. With a good knowledge of the dust
temperature and the mass absorption coefficient in the 21$\mum$
wavelength range of the carrier, one can therefore estimate the
abundance of a typical element locked up in dust required to emit
the observed power $\Etot$  in the 21$\mum$ band
\begin{equation}
\xstar = \frac{\nx\,\Etot} {\mud\,\mHstar \int_{21\mum\,{\rm band}}
\kabs(\lambda)\times 4\pi B_{\lambda}(\Td)d\lambda} ~.
\label{equation:X-H}
\end{equation}

It can be seen that for a given $\Etot$, the required abundance
$\xstar$ is inversely proportional to the integral of absorption
coefficient $\kabs$ and the dust radiation intensity
$B_{\lambda}(\Td)$ over the 21$\mum$ band range.

We approximate the mass absorption coefficient profile
with a Drude function
\begin{equation}
\kabs(\lambda) = \frac{\kabsint\,2\gamma/\pi}
{\left(\lambda-\lambdap^2/\lambda\right)^2+\gamma^2}~,
\label{equation:drude}
\end{equation}
where $\lambdap\approx 20.1\mum$
and $\gamma\approx 2.2\mum$ are respectively
the peak wavelength and FWHM of the profile;\footnote{%
  The Drude profile, closely resembling a Lorentz profile
  and having more extended wings than a Gaussian profile,
  is expected for classical damped harmonic oscillators
  \citep{Li2008Optical}. With $\lambdap$\,=\,20.1$\mum$
  and $\gamma$\,=\,2.2$\mum$, the Drude profile fits well
  the observed 21$\mum$ emission feature except the blue-wing
  (which is a bit more extended than that observed),
  while the Gaussian profile is a bit too narrow in the red-wing.
  To be more physical, we should fit the observed 21$\mum$
  emission feature with $\sum_j\kabs(\lambda)\,B_\lambda(T_j)$,
  a sum of products of a Drude mass absorption profile
  and Planck functions of a range of temperatures
  (since the carrier of the 21$\mum$ feature is expected to have
  a range of thermal equilibrium temperatures
  in the 21$\mum$-feature-emitting shell).
  Indeed, we find that with $\lambdap$\,=\,20.1$\mum$,
  $\gamma$\,=\,1.85$\mum$, and $T\approx 90\K$,
  the product of a Drude profile and a single-temperature
  blackbody $\kabs(\lambda)\,B_\lambda(T)$ closely fits
  the observed 21$\mum$ emission feature
    (see the inserted panel in Fig.\,1h).
  In this work, we take $\lambdap$\,=\,20.1$\mum$
  and $\gamma$\,=\,2.2$\mum$ in order to maximize the energy
  output in the 21$\mum$ feature so as to minimize the dust
  abundance requirement.
     }
and
$\kabsint \equiv \int_{21\mum\,{\rm band}} \kabs(\lambda) d\lambda$
is the mass absorption coefficient integrated over the 21$\mum$
band. The adoption of a Drude function and the values of $\lambdap$
and $\gamma$ which match the observed feature profile mostly is the
best condition for the absorption coefficient. The abundance under
such condition is the minimum. By substituting
Eq.\,(\ref{equation:drude}) into Eq.\,(\ref{equation:X-H}), the
lower limit of $\xstar$ becomes
\begin{equation}
\xstar_{\rm min} =\frac{\nx\,\Etot} {8\mud\mHstar\,\gamma\,\kabsint
\int_{21\mum\,{\rm band}}
\frac{B_{\lambda}(\Td)}{\left(\lambda-\lambdap^2/\lambda\right)^2+\gamma^2}
d\lambda} ~. \label{eq:XtoHmin}
\end{equation}
%
%
Note that the integration is continuum-subtracted.
Since the Planck function $B_\lambda(T)$ is
a monotonically increasing function of $\Td$,
the minimum abundance of element X, $\xstar_{\rm min}$,
becomes a decreasing function of $\Td$ and $\kabsint$.
For a fixed $\Etot$, given a reasonable dust temperature range
(usually $\sim$\,100--200K for the circumstellar envelope
of post-AGB stars) and $\kabsint$ from laboratory data,
we can estimate the range of $\xstar_{\rm min}$.
For a viable candidate, the required minimum
abundance $\xstar_{\rm min}$ must not exceed
what is available.

If $\kabs(\lambda)$ is measured in the laboratory, $\kabsint$ can be
obtained from the direct integration of the measured profile of
$\kabs(\lambda)$. In a few cases, $\kabs(\lambda)$ is not directly
measured, we obtain $\kabsint$ from the absorption cross section
$C_{\rm abs}$ or the absorption efficiency $Q_{\rm abs}$ calculated
through Mie theory
\begin{eqnarray}
\kabsint &=& \int_{21\mum\,{\rm band}} \kabs(\lambda) d\lambda
\nonumber
\\
&=&\int_{21\mum\,{\rm band}}\frac{C_{\rm
abs}(a,\lambda)}{M}d\lambda \nonumber \\
&=&\frac{3}{4\rho_{\rm
d}}\int_{21\mum\,{\rm band}}
  \frac{Q_{\rm abs}(a,\lambda)}{a}d\lambda ~,  \label{eq:transfrom}
\end{eqnarray}
where $M$ and $a$ are respectively the mass and radius of the dust
grain (we assume that the dust is spherical), and $\rho_{\rm d}$ is
the mass density of the dust.

\section{The Tester: HD 56126 }
A successful candidate carrier should be able to explain the
observed 21$\mum$ feature in all sources. A failure in a single
source would be sufficient to rule out the candidate. To examine
whether the carriers can account for the observed feature strength,
we choose HD\,56126, a prototypical 21$\mum$ feature source, as the
tester.

HD\,56126 (IRAS\,07134+1005) is selected for the following reasons:
(1) It is the strongest 21$\mum$ feature emitter and its emission
flux was accurately measured; (2) It is one of the best studied
21$\mum$ feature sources and its basic parameters are accurately
determined. \citet{Hony...2003A&A...402..211H} built a detailed dust
radiative transfer model for the circumstellar envelope and derived
the composition and mass of the dust shell of HD\,56126.
\citet{VanWinckel2000A&A...354} performed a homogeneous photospheric
abundance analysis of HD\,56126 (as well as several other 21$\mum$
feature sources). We summarize in Table \ref{tab:basicdata} the key
relevant stellar and circumstellar parameters which will be used in
later analysis. The parameter $M_{\rm H}$ is not very certain in the
range $\sim$\,0.16--0.44\,$M_\odot$. We generously take the high end
($M_{\rm H}\approx 0.44\,M_\odot$) for our analysis. Since the
required minimum abundance $\xstar_{\rm min}$ is inversely
proportional to $M_{\rm H}$ (see eq.\,\ref{eq:XtoHmin}), if a model
based on the high end $M_{\rm H}\approx 0.44\,M_\odot$ already
exceeds the elemental budget it should certainly be rejected.

By assuming that the X atoms available in the circumstellar
envelopes around the 21$\mum$ feature sources
(i.e. $\xstar_\star$) are {\it all} depleted in the dust
species proposed as a carrier candidate,
we obtain the {\it maximum} dust mass (relative to H)
of the dust species containing the key element X
from Eq.\,\ref{eq:XtoH}
\begin{equation}
\mdmax = \mud\,\xstar_\star/\nx ~.
\label{eq:Mdmax}
\end{equation}
This is the {\it upper} limit on the amount of dust containing
element X {\it available} to account for the 21$\mum$ feature.
For a given dust temperature $\Td$,
from Eq.\,\ref{eq:XtoHmin} we obtain the {\it lower} limit on
the amount of dust (relative to H)
{\it required} to account for the 21$\mum$ feature
\begin{eqnarray}
\mdmin &=& \mud\,\xstar_{\rm min}/\nx \nonumber \\
&=&\frac{\Etot}{8\gamma\,\kabsint
  \int_{21\mum\,{\rm band}}\frac{B_{\lambda}(\Td)}
  {\left(\lambda-\lambdap^2/\lambda\right)^2+\gamma^2}
d\lambda} ~. \label{eq:Mdmin}
\end{eqnarray}
Apparently, for a valid carrier candidate,
the minimum required mass $\mdmin$ should not
exceed the maximum available mass $\mdmax$.

\begin{figure*}
\includegraphics[width=16cm]{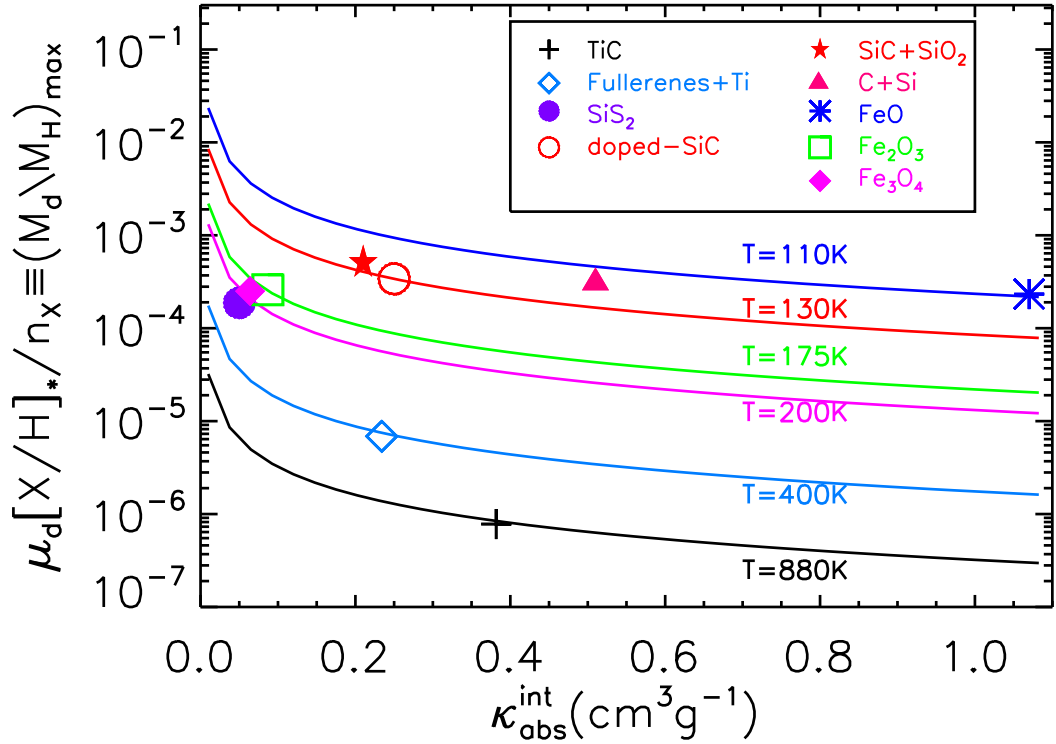}\\
\caption{\label{fig:X2Hmin}
         $\kabsint$, the integrated mass
         absorption coefficient
         of the 21$\mum$ band (with the continuum underneath
         the band subtracted)
         versus $\left(M_{\rm d}/M_{\rm H}\right)_{\rm max}
          \equiv \mu_{\rm d}\,\xstar_\star/\nx$,
         the maximum dust mass (relative to H) of the species
         containing the key element ``X''
         (obtained by assuming that all X atoms are tied in
          that particular dust species), as denoted by
          cross (TiC nanoclusters),
          open diamond (fullerenes coordinated with Ti atoms),
          filled circle (SiS$_2$),
          open circle (doped-SiC),
          filled star (SiO$_2$-coated SiC),
          filled triangle (carbon-silicon mixtures),
          asterisk (FeO),
          open square (Fe$_2$O$_3$), and
          filled diamond (Fe$_3$O$_4$).
          Also plotted (as solid lines) are the minimum dust mass
          required to account for the emitted power
          of the 21$\mum$ feature
          $\left(M_{\rm d}/M_{\rm H}\right)_{\rm min}
          \equiv \mu_{\rm d}\,\xstar_{\rm min}/\nx$
          as a function of $\kabsint$
          for a range of temperatures.}
\end{figure*}

In Fig.\,\ref{fig:X2Hmin} we plot $\mdmax$ against
$\kabsint$ for the proposed carrier candidates.
We also plot $\mdmin$ as a function of $\kabsint$
for a range of temperatures $\Td$.
Note that the temperature at which
$\mdmax \approx \mdmin$ is the {\it lowest}
temperature the dust should have.
For example, at $T\approx 880\K$ TiC nanoclusters
have $\mdmax \approx \mdmin$. In order for
$\mdmin$ not exceeding $\mdmax$, TiC dust must
have $\Td > 880\K$, otherwise there is simply
not enough dust material.

\section{Assessing Individual Carrier Candidates}\label{sec:candidate}
%
We assess proposed individual carrier candidates by examining (1)
whether they are capable of emitting
    the observed large amount of energy
    in the 21$\mum$ band
    without requiring more dust material
    than available, and
(2) whether the candidate carrier produces
    (undetected) secondary features.
%
In view of the crucial role of elemental abundances in this
assessment, we divide the carrier candidates into four groups:
titanium-bearing, sulfur-bearing, silicon-bearing and iron-bearing
grains.


 The 3.3, 7.7 and 11.3$\mum$ PAH features and the much more
prominent 30$\mum$ feature which is generally attributed to MgS dust
\citep{Goebel1985ApJ...290L..35G, Hony2002A&A...390..533H} are seen
in fourteen of the sixteen known 21$\mum$ feature sources
(\citealt{Kwok1999IAUS..191}, \citealt{Hrivnak2009arXiv0902.0077H}).
Except these features, the 21$\mum$ sources do not universally show
any additional dust features.
However, in addition to the 21$\mum$ feature,
some of the proposed carrier candidates exhibit
strong spectral features at other wavelengths as well
which are not detected at all or very weak
in the 21$\mum$ sources.

In view  of the wide detection of the PAH features (and some flat
plateau attributed to HAC) and the 30$\mum$ feature in the 21$\mum$
sources, \citet{Buss1993ApJ...415} suggested that the 21$\mum$
feature might originate from some organic molecules like PAHs, while
\citet{Goebel1993A&A...278} argued that sulfide could be the carrier
candidate. But the proposition of a carrier candidate based on its
matching to these accompanying features is potentially problematic
since they vary from source to source, while the profile of the
21$\mum$ feature is rather universal (i.e. with the same peak
wavelength, a very similar FWHM and asymmetrical shape with a long
red tail) indicating that the carrier of this feature should be the
same in different sources. We also note that observationally, there
does not appear to show any correlation between the strength of the
21$\mum$ feature with that of the 30$\mum$ feature
\citep{JZL2009EPS}.

For a given proposed carrier, the predicted intensity ratios of the
associated features to the primary 21$\mum$ feature must either be
compatible with that observed (in case of detection) or not exceed
the upper limit (in case of non-detection). SiC dust with carbon
impurities, once considered as a promising candidate due to its
abundant occurrence \citep{Speck2004ApJ...600}, was challenged based
on the much higher model-predicted intensity ratio of the 11.3$\mum$
feature to the 21$\mum$ feature than observed
\citep{Jiang2005ApJ...630L}.

\subsection{Titanium-Bearing Grains\label{sec:Ti}}
Titanium is a relatively rare element in the solar system
([Ti/H]$_\odot$\,$\approx$\,9.77$\times10^{-8}$,
\citealt{Grevesse1989AIPC..183....9G}),
and the Ti abundance in the metal-poor star HD\,56126
is even lower:
[Ti/H]$\approx$1.3$\times$10$^{-8}$ \citep{VanWinckel2000A&A...354}.
Although some Ti-bearing grains
(e.g. TiC nanoclusters and fullerenes
      coordinated with Ti atoms)
have a 21$\mum$ feature with a close similarity
to the 21$\mum$ feature observed in PPNe,
the deficiency of titanium would be a potential problem.

\subsubsection{Titanium Carbide}
\citet{vonHelden...2000Sci...288..313V} argued that the 21$\mum$
feature may arise from titanium carbide nanoclusters (made of 27 to
125 atoms). The laboratory-measured spectral profiles of TiC
nanoclusters match almost perfectly with the intrinsic profile of
the observed 21$\mum$ feature, better than any other candidate
materials, although bulk TiC dust does not display a strong 21$\mum$
band (see \citealt{Henning2001AcSpe..57..815H,
Kimura2003MNRAS.343..385K}).
The TiC hypothesis gains its strength from the
identification of presolar TiC grains (with radii $\sim$100{\AA}) in
primitive meteorites as nanometer-sized inclusions embedded in
micrometer-sized presolar graphite grains
\citep{Bernatowicz1996ApJ...472..760B}. However, since Ti is a rare
element, the abundance test would be a neck to the TiC hypothesis.
Indeed, three different groups have already challenged
the TiC hypothesis from the Ti abundance point of view
\citep{Hony...2003A&A...402..211H,
Chigai...2003ApJ...587..771C,
Li...2003ApJ...599L..45L}.
They all pointed out that there may not be enough titanium
to account for the observed strength of this feature.
Here we add another piece of evidence against the TiC hypothesis by
confronting it with the band strength constraint discussed in
\S\ref{sec:constraints}. Unlike previous studies, in this approach
we do not need to know the ultraviolet/optical absorption properties
of TiC. The experimental IR absorption spectrum of TiC nanoclusters
was measured \citep{vonHelden...2000Sci...288..313V} and fitted with
a Lorentz oscillator model \citep{Chigai...2003ApJ...587..771C}. The
observational emission spectrum of HD\,56126 and the experimental
Lorentz profile of the 21$\mum$ band are shown in
Fig.\,\ref{fig:kabs_21um}a.

The integrated mass absorption coefficient
of nano TiC is $\kabsint \approx 0.38\cm^3\g^{-1}$
for the 21$\mum$ band, as derived from the experimental
spectrum of \cite{vonHelden...2000Sci...288..313V}.
The emission temperature associated with the 21$\mum$
carrier is not well constrained.

Exposed to the stellar radiation, TiC nanocrystals,
because of their small heat capacities, will not attain
an equilibrium temperature; instead, they will be
transiently heated by single photons
(see \citealt{Draine_Li2001ApJ...551..807}).
The stellar atmospheric spectrum of HD\,56126
peaks at $\lambda \approx 0.41\mum$, i.e., a typical
stellar photon has an energy of
$\langle h\nu \rangle \approx 3\eV$.
With a Debye temperature of $\Theta \approx 614\K$
\citep{Pierson1996}, when heated by a 3$\eV$ photon,
for a TiC nano cluster as small as Ti$_{14}$C$_{13}$
(consisting of $3\times 3\times3$ atoms)
the peak temperature is only $T_{\rm peak}\approx 268\K$.

We calculate from Eq.\,(\ref{eq:XtoHmin})
the minimum abundance requirement of Ti$_{14}$C$_{13}$
dust to be $\Timin \approx 1.97\times10^{-7}$
at $\Td = 268\K$. Since the titanium abundance in HD\,56126
is measured to be $\Tistar \approx 1.3\times10^{-8}$
(see Table \ref{tab:basicdata}),
the minimum abundance requirement
(at $\Td = 268\K$) exceeds the available abundance
by a factor of $\simali$15.

For a given exciting photon energy
(say, $h\nu = 3\eV$),
the peak temperature decreases with the nanocluster size:
$T_{\rm peak}\approx 216\K$ for Ti$_{32}$C$_{32}$
(consisting of $4\times 4\times4$ atoms)
and $T_{\rm peak}\approx 183\K$ for Ti$_{72}$C$_{53}$
(consisting of $5\times 5\times5$ atoms).
Therefore, larger TiC nanoclusters would require
even more Ti atoms to account for the observed
21$\mum$ intensity.

In fact, based on the available Ti abundance in HD\,56126,
nano TiC grains have to reach a temperature at least
as high as $\sim$\,880\,K to achieve the observed emission
strength [i.e. $\mdmin \approx \mdmax$; see Fig.\,\ref{fig:X2Hmin}],
while the peak temperature is only $\sim$\,268\,K even
for a small cluster consisting of only 27 atoms.
Apparently, the TiC hypothesis requires too much Ti
to be viable, even under the most optimal condition
that all Ti atoms are locked in TiC nanocrystals.

\subsubsection{Fullerenes Coordinated with Ti Atoms}
\citet{Kimura2005ApJ...632} found that the laboratory spectra of
large-cage carbon particles (fullerenes) coordinated with Ti atoms
have a characteristic feature at $\simali$20.3$\mum$ closely
resembling the 21$\mum$ feature of post-AGB stars
(see Fig.\,\ref{fig:kabs_21um}b).
They attributed the origin of the 21$\mum$ feature
to the vibrational interaction between Ti atoms and
fullerene cages.
In order to obtain the integrated mass absorption
coefficient $\kabsint$ only over the 21$\mum$ band,
the continuum absorption needs to be subtracted.
Based on the continuum spectrum between 15--18.5$\mum$
and 22--24$\mum$, we use a two order polynomial
to fit the continuum underneath the 21$\mum$ feature.
In the following sections, unless otherwise stated,
the continuum absorption is determined in the same way.
The resulting continuum-subtracted,
integrated mass absorption coefficient
for the 21$\mum$ is $\kabsint\approx 0.233\cm^3\g^{-1}$.

Similar to TiC nanoclusters, fullerenes will also subject
to stochastic heating in the circumstellar envelope around
HD\,56126. With a Debye temperature of
$\Theta \approx 185\K$,\footnote{%
  See {\sf http://www.sesres.com/PhysicalProperties.asp}.
  }
when heated by a typical photon of 3$\eV$ in HD\,56126,
even C$_{60}$ is only heated to a peak temperature
of $T_{\rm peak}\approx 89\K$.\footnote{%
  For larger fullerenes, we expect a lower peak
       temperature $T_{\rm peak}$ and therefore require
       a higher minimum abundance $\Timin$,
       indicating a more severe Ti abundance budget shortage.}

With $\kabsint\approx 0.233\cm^3\g^{-1}$
and $T_{\rm peak}\approx 89\K$,
from Eq.\,(\ref{eq:XtoHmin}) we estimate the minimum
abundance requirement of fullerenes with Ti atoms
to be $\Timin \approx 8.04\times10^{-6}$,
exceeding the available Ti abundance
$\Tistar \approx 1.3\times10^{-8}$
by a factor of $\simali$618.
In order not to violate the abundance constraint
(i.e. $\Timin < \Tistar$
or $\mdmin < \mdmax$),
the dust temperature should be higher than 400\,K
(see Fig.\,\ref{fig:X2Hmin}), while the peak
temperature of the C$_{60}$\,+\,Ti dust is only $\simali$89\,K.
Therefore, large fullerenes coordinated with Ti atoms
are unlikely the carrier of the 21$\mum$ feature
seen in PPNe.

\subsection{Sulfur-Bearing Grains:
            Silicon Disulfide \label{sec:sis2}}
The formation of sulfide is very likely to occur
in carbon-rich circumstellar environments.
As early as thirty years ago,
\citet{Lattimer1978ApJ...219..230L}
predicted the possible presence of various
sulfur-bearing materials in carbon-rich systems.
For several kinds of predicted sulfides,
including MgS, FeS, SiS$_2$ and CaS,
the absorption spectra and coefficients
have been measured in laboratory
\citep{Nuth1985ApJ...290L}.
The prominent 30$\mum$ dust feature seen
in AGB, post-AGB and PNe, amounting to more than 20\%
of the total IR flux, is attributed to MgS
\citep{Goebel1985ApJ...290L..35G, Hony2002A&A...390..533H}.
Since most of the 21$\mum$ sources also exhibit
a strong feature at 30$\mum$,
it is not unreasonable to postulate that some
sulfur-bearing grains may also contribute to the 21$\mum$ feature.
Indeed, the laboratory spectrum of SiS$_2$ displays
a prominent feature at $\simali$22$\mum$
\citep{Nuth1985ApJ...290L}. \citet{Goebel1993A&A...278}
further suggested SiS$_2$ solids as the
material responsible for the 21$\mum$ feature.

Based on the laboratory spectra of \citet{Goebel1993A&A...278}
and \citet{Nuth1985ApJ...290L},
we obtain $\kabsint\approx 0.05\cm^3\g^{-1}$
for the 21$\mum$ band of SiS$_2$ after subtracting
the continuum (see Fig.\,\ref{fig:kabs_21um}).
%
%
In addition to the 21$\mum$ band, the laboratory absorption spectrum
of SiS$_2$ also exhibits a secondary feature at 17$\mum$ that is
never observed in the 21$\mum$ sources \citep{Kraus1997A&A...328}.
In order to sufficiently suppress the 17$\mum$ feature so that it
remains un-noticeable, the dust temperature needs to be $<$\,100\,K.
For $\Td = 100\K$ we estimate the minimum abundance requirement of
SiS$_2$ to be $\Smin \approx 9.59\times10^{-5}$, exceeding the
available S abundance $\Sstar \approx 4.07\times10^{-6}$ by a factor
of $\simali$24. As shown in Fig.\,\ref{fig:X2Hmin}, in order for the
SiS$_2$ model to satisfy the abundance constraint (i.e. $\Smin
<\Sstar$ or $\mdmin < \mdmax$), SiS$_2$ dust needs to be hotter than
$\simali$200\,K.

\citet{Posch2004ApJ...616} also recognized that very low dust
temperatures (significantly lower than 100\,K) would be required to
make the secondary feature at 17$\mum$ negligible in strength
compared to the 21$\mum$ SiS$_2$ feature. Although it is not
possible to calculate the equilibrium temperature of SiS$_2$ due to
the lack of knowledge of its visual and near-IR optical constants,
\citet{Posch2004ApJ...616} argued that ``it is hardly conceivable
that SiS$_2$ is so transparent in the visual range as to remain much
colder than 100\,K''. In our approach we actually do not need to
know the exact temperature of SiS$_2$: (1) if $T<100\K$ -- although
the 17$\mum$ feature will be suppressed, one requires too much S;
(2) if $T>200\K$ -- although there will not be a S budget problem,
the 17$\mum$ feature will be prominent. Therefore, our approach
readily ruled out SiS$_2$.

Moreover, it is worth noting that sulphur may not be completely
locked in SiS$_2$. The prominent 30$\mum$ feature, if indeed
arising from MgS, would consume a significant portion of
the sulphur available in the circumstellar envelopes
around the 21$\mum$ sources.
According to \citet{Zhukovska2008A&A...486..229Z},
MgS has priority in the cooling sequence of
sulphur-bearing solid compound.

\subsection{Silicon-Bearing Grains\label{sec:Si}}
With four valence electrons, Si  easily reacts with other atoms such
as C, O, Fe and Mg to form various types of silicates. In the O-rich
dust shells of evolved stars (from AGB stars to PNe), both amorphous
and crystalline silicates are detected through their numerous
emission bands. Over 4000 sources are detected to have the most
common silicate features at 9.7$\mum$ and 18$\mum$. While in the
C-rich dust shells like those of the 21$\mum$ sources, relatively
simple Si-bearing compounds are formed (e.g. SiC, SiO$_2$, and
SiS$_2$). The broad 11.3$\mum$ feature seen in five of the sixteen
21$\mum$ sources \citep{Kwok1999IAUS..191} is identified to arise
from the Si--C stretching mode of SiC. Being chemically active and
abundant in circumstellar envelopes, several Si-bearing dust species
have been proposed to be the carrier of the 21$\mum$ feature.

Silicon is an abundant element in the universe
(its solar abundance is $\SitoH_\odot \approx 3.55\times10^{-5}$,
\citealt{Grevesse1989AIPC..183....9G}),
about one order of magnitude higher than sulphur.
Unfortunately, the Si abundance $\Sistar$ has not been measured for
our tester HD\,56126. We take the following approach to roughly
estimate $\Sistar$ of HD\,56126: (1) We compile the abundance data
of all 21$\mum$ sources
    and find that five 21$\mum$ sources
    -- IRAS\,04296, IRAS\,22223, IRAS\,23304,
        (\citealt{VanWinckel2000A&A...354}),
       and IRAS\,05113, IRAS\,22272 (\citealt{Reddy2002ApJ...564..482R}) --
       have known Si abundance.
(2) We calculate the abundance ratios of Si to Ca
    and of Si to S for all five sources
    and find that these ratios do not vary much.\footnote{%
This may not be unexpected since S, Si, and Ca
      all are $\alpha$ elements.
      Let $\left\{{\rm X/H}\right\}\equiv \log_{10}\XtoH\,+\,12$
      and $\left\{{\rm X/Y}\right\} \equiv
          \left\{{\rm X/H}\right\}\,-\,\left\{{\rm Y/H}\right\}$.
      For IRAS\,04296, IRAS\,22223, IRAS\,23304,
      IRAS\,05113, and IRAS\,22272,
      the Si to Ca abundance ratios are respectively
      $\left\{{\rm Si/Ca}\right\}$\,$\approx$\,1.80, 1.57,
      1.69, 1.45, and 1.49, with a mean ratio of
      $\langle\left\{{\rm Si/Ca}\right\}\rangle$\,$\approx$\,1.55
      (and a standard deviation of $\sigma\approx 0.12$).
      The Si to S abundance ratios vary a bit more compared to
      $\left\{{\rm Si/Ca}\right\}$; they are respectively
      $\left\{{\rm Si/S}\right\}$\,$\approx$\,0.70, 0.59,
      0.57, 0.22, and 0.28 for IRAS\,04296, IRAS\,22223,
      IRAS\,23304, IRAS\,05113, and IRAS\,22272,
      with a mean ratio of
      $\langle\left\{{\rm Si/S}\right\}\rangle$\,$\approx$\,0.42
      (and a standard deviation of $\sigma\approx 0.18$).
      }

(3) By assuming that the Si to Ca and Si to S abundance ratios
    of HD\,56125 are similar to that of the other five 21$\mum$
    sources, we estimate $\Sistar\approx 6.31\ppm$
    for HD\,56126 from $\left\{{\rm Si/Ca}\right\}\approx 1.55$
    and $\Castar\approx 0.18\ppm$ (\citealt{VanWinckel2000A&A...354}),
    or $\Sistar\approx 10.7\ppm$
    from $\left\{{\rm Si/S}\right\}\approx 0.42$
    and $\Sstar\approx 4.07\ppm$.
(4) We finally take the Si abundance of HD\,56126 to be
    the average of that estimated from Ca and S,
    $\Sistar\approx 8.5\ppm$.

\subsubsection{Doped-SiC\label{sec:doped-SiC}}
Silicon carbide (SiC) grains with impurities were suggested to be
the carrier of the 21$\mum$ feature, based on laboratory data that
doped SiC grains exhibit a resonance at $\sim$ 21$\mum$
\citep{Speck2004ApJ...600}. This proposal gains strength from the
fact that SiC is a common dust species in carbon-rich circumstellar
envelopes. The presence of circumstellar SiC grains was first
revealed by \citet{Gilra1972PhDT}. Nowadays, SiC is believed to be
the contributor of the well-known IR dust feature at
11.3$\mum$ observed in more than 700 carbon stars
\citep{Kwok...1997ApJS..112..557K}. Since five of the sixteen
21$\mum$ sources (i.e. about 40\%) exhibit the 11.3$\mum$ feature,
it is not unreasonable to consider both features coming from the
same carrier. In particular, the distributions of the dust emitting
at 21$\mum$ and 11.3$\mum$ are co-spatial in two 21$\mum$ sources
(HD\,56126 and IRAS\,Z02229+6208; \citealt{Kwok2002ApJ...573}),
which also supports the SiC hypothesis as the 21$\mum$ carrier.

However, the 21$\mum$ feature is {\it secondary} and much weaker
than the {\it primary} 11.3$\mum$ feature in the experimental
spectrum of SiC \citep{Speck2004ApJ...600}.
\citet{Jiang2005ApJ...630L} assumed a single Lorentz oscillator
model for the 21$\mum$ feature with $Q_{\rm abs}(21\mum)/a$ treated
as a free parameter (where $Q_{\rm abs}(\lambda)$ is the absorption
efficiency at wavelength $\lambda$ and $a$ is the dust size). They
found that the predicted 11.3$\mum$ feature is still much stronger
than observed even with $Q_{\rm abs}(21\mum)/a$\,=\,$10^4\cm^{-1}$.
It is unlikely to expect the strength of the 21$\mum$ feature, a
secondary feature of SiC caused by impurities, would be much
stronger than that of the primary 11.3$\mum$ feature for which
$Q_{\rm abs}(11.3\mum)/a\approx1.5\times10^4\cm^{-4}$
\citep{Jiang2005ApJ...630L}. Therefore, doped-SiC cannot produce the
21$\mum$ feature without producing too strong a 11.3$\mum$ feature
to be consistent with observations.

\citet{Jiang2005ApJ...630L} calculated the equilibrium
temperatures and $\kabs$ of doped-SiC of different grain sizes
and a range of $Q_{\rm abs}(21\mum)/a$ values.
For $\alpha$-SiC, the equilibrium temperature is in a narrow range
($\simali$60--80\,K), insensitive to the assumed $Q_{\rm
abs}(21\mum)/a$ values (see Fig.\,2 of
\citealt{Jiang2005ApJ...630L}). We see that even with $Q_{\rm
abs}(21\mum)/a$\,=\,$10^{4}\cm^{-1}$ (which implies $\kabsint\approx
0.25\cm^3\g^{-1}$), the required Si abundance is at least
$\simali$$8.76\times 10^{-4}$ (for $T\approx 80\K$) and
$\simali$$2.31\times 10^{-3}$ (for $T\approx 60\K$), way too much
compared with $\Sistar\approx 8.5\ppm$ estimated for HD\,56126.

Unlike $\alpha$-SiC, the equilibrium temperatures
of $\beta$-SiC are sensitive to $Q_{\rm abs}(21\mum)/a$
and become much higher for smaller $Q_{\rm abs}(21\mum)/a$
(see Fig.\,3 of \citealt{Jiang2005ApJ...630L}).
We will consider cases with both large and small
$Q_{\rm abs}(21\mum)/a$.
For $Q_{\rm abs}(21\mum)/a = 100\cm^{-1}$
(i.e. $\kabsint\approx 0.003\cm^3\g^{-1}$),
submicron-sized $\beta$-SiC dust has
an equilibrium temperature $\simali$100\,K
(see Fig.\,3 of \citealt{Jiang2005ApJ...630L}),
suggesting that one requires at least
$\SitoH_{\rm min}$\,$\approx$\,$3.33\times 10^{-3}$.
For $Q_{\rm abs}(21\mum)/a = 10^4\cm^{-1}$
(i.e. $\kabsint\approx 0.28\cm^3\g^{-1}$),
with a typical equilibrium temperature $\simali$60\,K
(see Fig.\,3 of \citealt{Jiang2005ApJ...630L}),
the minimum required Si abundance is
$\SitoH_{\rm min}$\,$\approx$\,$3.80\times 10^{-3}$.
Therefore, for both $\alpha$-SiC and $\beta$-SiC, one needs way too
much Si, exceeding the Si abundance available in HD\,56126
($\simali$8.50$\ppm$) by a factor of two magnitudes. So it is secure
to conclude that doped-SiC can not be the carrier of the 21$\mum$
feature.

\subsubsection{SiC Core-SiO$_2$ Mantle Grains
               \label{sec:core-mantle}}
\citet{Clement2003ApJ...594} carried out
laboratory experiments and found that pure
SiC nano-particles can be quite easily oxidized
at their surfaces to form SiO$_2$-coated SiC grains.
The oxidization can even reach a considerable volume
fraction of the particles.
The laboratory spectra of the partially oxidized SiC
grains show an absorption feature at $\simali$21--22$\mum$.
\citet{Posch2004ApJ...616} suggested that dust composed of
a SiC core and a SiO$_2$ mantle may be the carrier
of the 21$\mum$ feature.

From the $Q_{\rm abs}(\lambda)/a$ values of
\citet{Posch2004ApJ...616}, we obtain
$\kabsint$\,$\approx$\,$0.21\cm^3\g^{-1}$.
The equilibrium temperatures of submicron-sized
SiC core-SiO$_2$ mantle grains is $\simali$104--126$\K$
in the 21$\mum$ emitting region of HD\,56126 \citep{Posch2004ApJ...616}.
The minimum requirement of Si is
$\simali$1.00$\times$10$^{-5}$--3.22$\times$10$^{-5}$,
which is slightly above the available Si abundance
of $\Sistar$\,$\approx$\,8.50$\ppm$.

However, these core-mantle grains have two prominent features
(see Fig.\,\ref{fig:kabs_21um}d)
at 8.3$\mum$ (arising from the SiO$_2$ mantle)
and 11.3$\mum$ (arising from the SiC core).
The 8.3$\mum$ feature is never seen in the 21$\mum$ sources
except three 21$\mum$ sources appear to have a plateau
at $\simali$8$\mum$ with an average FWHM of $\simali$4$\um$
which is more likely from HAC
or the C--C stretching mode of PAHs
\citep{Kwok2001ApJ...554L..87K}.
As far as the 11.3$\mum$ feature is concerned,
we compare the flux ratio of the 11.3$\mum$ band
to the 21$\mum$ band observed in HD\,56126
[$F(11.3\mum)/F(21\mum)<0.012$]\citep{Hony...2003A&A...402..211H}
with that predicted from the SiC core-SiO$_2$ mantle model.
It is found that the dust should be colder than $\simali$70$\K$ in
order not to produce too strong a 11.3$\mum$ feature. However, the
equilibrium temperature of submicron-sized SiC core-SiO$_2$ mantle
grains is $\simali$104--126$\K$ in the 21$\mum$ emitting region of
HD\,56126 \citep{Posch2004ApJ...616}. Moreover, if the dust
temperature is as low as $<$\,70\,K,

the minimum Si abundance requirement
would be $\simali$8.05$\times$10$^{-4}$,
exceeding the Si abundance available in HD\,56126
($\Sistar\approx 8.50\ppm$) by a factor of $\simali$100.

In addition, the opacity profile of these grains has
a shoulder in the red-wing of the 21$\mum$ band
which is not observed in the 21$\mum$ sources.
Therefore, both the secondary features and the 21$\mum$ profile
are discrepant with observations, suggesting that SiC core-SiO$_2$
mantle grains are not a valid carrier candidate.

\subsubsection{Solid-Solution Phase of Carbon and Silicon
               with a Diamond Structure}\label{sec:c-si mixture}
\citet{Kimura2005crystal...275} measured the IR absorption
spectra of silicon-containing carbon films prepared by ion
sputtering of carbon and silicon carbide mixture pellets. The
carbon-silicon mixture film, composed of a solid-solution phase of
carbon and silicon with a diamond structure, show significant
absorption features at 9.5$\mum$ and 21$\mum$
(see Fig.\,\ref{fig:kabs_21um}e).
Thus, \citet{Kimura2005crystal...275} suggested that the 21$\mum$
feature observed in PPNe may arise from
the solid-solution phase of carbon and silicon
with a diamond structure.

However, the 9.5$\mum$ feature is far stronger than
the 21$\mum$ feature in the carbon-silicon mixture film,
but it is never seen in the 21$\mum$ sources.
In order to suppress the 9.5$\mum$ feature to
such a level that the total flux emitted in
the 9.5$\mum$ band is less than, say 10\% of that
in the 21$\mum$ feature, the critical dust temperature,
depending on the silicon percentage,
needs to be $<$\,102\,K for C--30\% Si film
or $<$\,120\,K for C--10\% Si film.

Unfortunately, there lacks a full knowledge of
the optical properties of such carbon-silicon mixtures
to determine their equilibrium temperatures.
We adopt the dielectric functions of diamonds
which may be a reasonable approximation of
the carbon-silicon mixtures with a diamond
structure \citep{Kimura2003crystal...255}.
We find the temperature of submicron-sized diamonds
in the 21$\mum$-emitting region of HD\,56126 is
$\simali$148--182$\K$.\footnote{%
  The Si abundance is less of an issue.
  From the absorption profiles of \citet{Kimura2005crystal...275},
  we obtain $\kabsint \approx 0.32\cm^3\g^{-1}$ for C--50\% Si film,
  $0.51\cm^3\g^{-1}$ for C--30\% Si film, and
  $0.40\cm^3\g^{-1}$ for C--10\% Si film.
  If the dust temperature is respectively higher than
  $\simali$130\,K, $\simali$110\,K, and $\simali$100\,K
  for C--50\% Si film, C--30\% Si film, and C--10\% Si film,
  the minimum required Si abundance would not exceed
  that available in HD\,56126.
  }
Therefore, the carbon-silicon mixtures, if they are indeed
responsible for the 21$\mum$ feature, would produce too strong a
9.5$\mum$ emission feature to be consistent with that observed in
the 21$\mum$ feature sources.

\subsection{Iron-Bearing Grains\label{sec:Fe}}
\subsubsection{Fe$_2$O$_3$ and Fe$_3$O$_4$\label{sec:iron oxides}}
Iron oxides were first suggested as the carrier of the 21$\mum$
feature by \citet{Cox1990A&A...236L}. \citet{Cox1990A&A...236L}
reported the detection of a 21$\mum$ emission feature in the IRAS
LRS spectra of ten HII regions and associated it with the 21$\mum$
band seen in PPNe. From the strength and the ``universality'' of
this feature in HII regions, iron oxides $\gamma$-Fe$_2$O$_3$
(maghemite) and Fe$_3$O$_4$ (magnetite) were assigned to this band.
However, \citet{Oudmaijer1995A&A...295L} later re-analyzed the IRAS
LRS spectra of these sources and concluded that the claimed 21$\mum$
band was just an artifact. This was confirmed in late 1990s when
some of these HII regions were observed with ISO which show no
footprints of a broad 21 or 20$\mum$ band
\citep{Posch2004ApJ...616}. In fact, the laboratory absorption
spectra of iron oxides of \citet{Cox1990A&A...236L} were quite
different from the intrinsic profile of the 21$\mum$ feature of PPNe
which had not been well identified until
\citet{volk...1999ApJ...516L..99V}. But of course the spectral
profile match might be improved if one considers dust shape and
clustering effects. Therefore iron oxides are not readily ruled out
just based on the imperfect spectral match.

Despite that the 21$\mum$ feature found in HII regions turns
out to be an artifact, iron oxides composed of cosmically abundant
elements and exhibiting a strong feature around 20$\mum$ deserve
a detailed study as being the carrier of the 21$\mum$ feature
seen in PPNe. In addition to $\gamma$-Fe$_2$O$_3$, other forms of
Fe$_2$O$_3$ exhibit a feature around 21$\mum$ as well.
\citet{Koike...1981Ap&SS..79...77K} indicated hematite
($\alpha$-Fe$_2$O$_3$) has several peaks
at $\simali$9.2, 18, 21 and 30$\mum$ in its
absorption spectrum. Similarly, the absorption spectrum of
Fe$_3$O$_4$ shows two features at $\simali$17 and 25$\mum$.

\begin{figure*}
  \includegraphics[width=16cm]{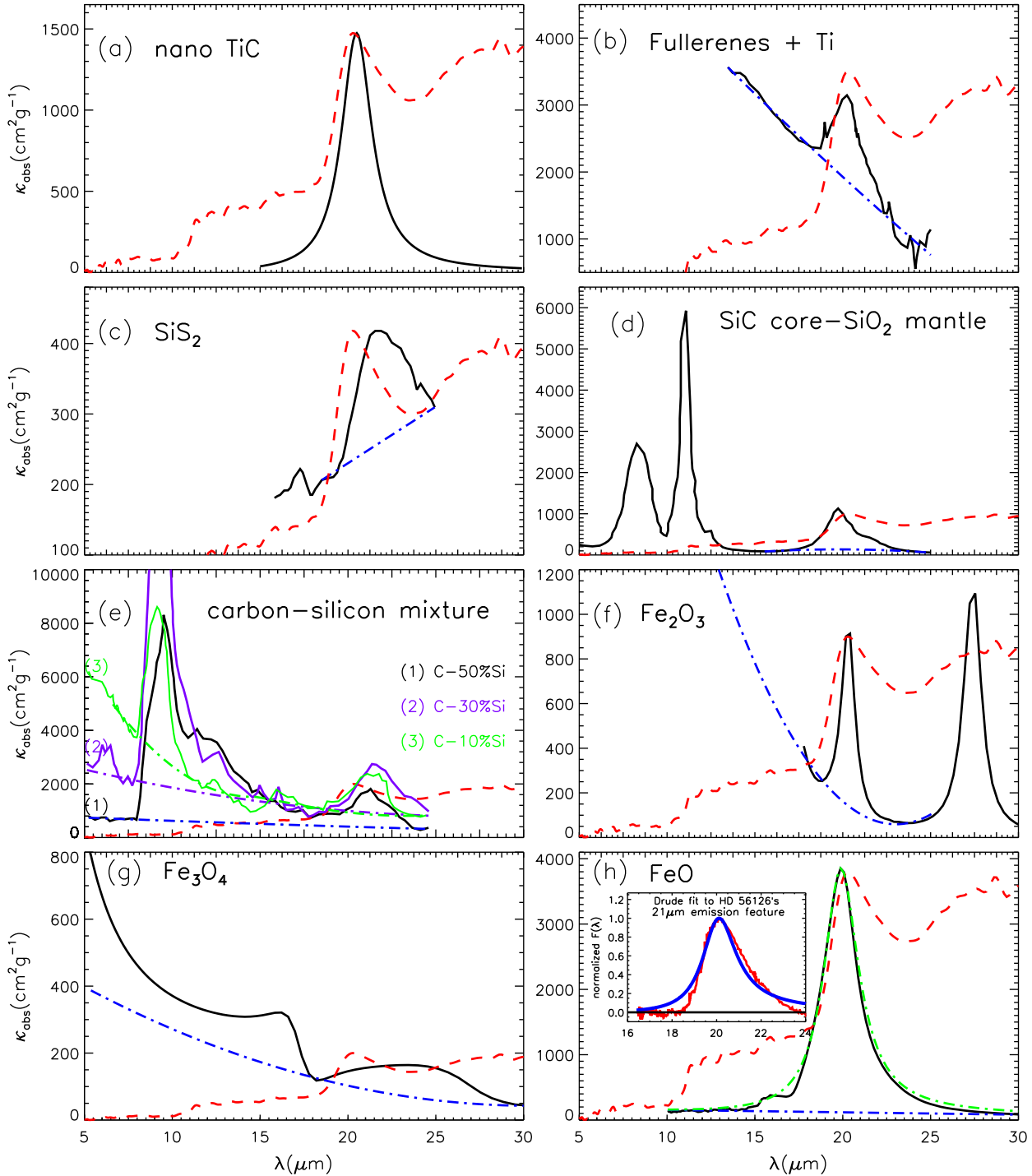}\\[8mm]
  \caption{Experimental or calculated mass absorption coefficient
           spectra $\kabs(\lambda)$ (black solid line) in the 21$\mum$
           wavelength range of eight carrier candidates for
           the 21$\mum$ feature seen in PPNe:
           (a) TiC nanoclusters,
           (b) large-cage carbon particles (fullerenes)
               coordinated with Ti atoms,
           (c) SiS$_2$ dust,
           (d) SiC core-SiO$_2$ mantle dust,
           (e) carbon and silicon mixtures,
           (f) Fe$_2$O$_3$,
           (g) Fe$_3$O$_4$, and
           (h) FeO.
           Also shown is the astronomical emission spectrum
           of HD\,56126 (scaled to match the $\kabs$ peak at 21$\mum$;
           red dashed line),
           a proto-typical 21$\mum$ feature source.
           The blue dot-dashed line plots the continuum
           underneath the experimental (or calculated) 21$\mum$
           mass absorption spectrum.
            Also shown in (h) is the Drude fit
           (with $\lambdap$\,=\,20.1$\mum$
            and $\gamma$\,=\,2.4$\mum$; green dot-dashed line)
           to the mass absorption profile $\kabs(\lambda)$ of FeO
           calculated from Mie theory
           (solid black line).
           The inserted panel in (h) fits
           the normalized, continuum-subtracted 21$\mum$ emission
           feature of HD\,56126 (solid red line)
           by $\kabs(\lambda)\,B_\lambda(T)$,
           the product of a Drude mass absorption profile
           (with $\lambdap$\,=\,20.1$\mum$ and
            $\gamma$\,=\,1.85$\mum$)
           and a blackbody of $T$\,=\,90\,K
           (solid blue line; see Footnote-1).
            }

\label{fig:kabs_21um}
\end{figure*}

\citet{Cox1990A&A...236L} only presented the normalized absorption
profiles of two iron oxides ($\gamma$-Fe$_2$O$_3$ and Fe$_3$O$_4$),
which cannot be used to calculate $\kabsint$. Based on the
unpublished optical constants of Fe$_2$O$_3$ and Fe$_3$O$_4$ from
the Jena
group\footnote{http://www.astro.uni-jena.de/Laboratory/OCDB/oxsul.html\#B1},
$\kabs(\lambda)$ and $\kabsint$ are calculated from Mie theory and
assuming a spherical radius of 0.1$\mum$. We obtain $\kabsint
\approx 0.09\cm^3\g^{-1}$ for Fe$_2$O$_3$ and $0.07\cm^3\g^{-1}$ for
Fe$_3$O$_4$. As shown in Fig.\,\ref{fig:kabs_21um}f and
Fig.\,\ref{fig:kabs_21um}g, Fe$_2$O$_3$ displays two strong bands at
$\simali$20.5$\mum$ and 27.5$\mum$, while
Fe$_3$O$_4$ exhibits two broad features at $\simali$16.5$\mum$
and 24$\mum$. Apparently, Fe$_3$O$_4$ does not fit
the 21$\mum$ feature seen in PPNe -- the model feature peak
is neither strong nor at the right wavelength
(see Fig.\,\ref{fig:kabs_21um}g).
For Fe$_2$O$_3$, even with temperature fine-tuning,
it is very hard to suppress the 27.5$\mum$ feature which
is not seen in the 21$\um$ sources.

Moreover, Fe$_2$O$_3$ and Fe$_3$O$_4$ may not be able
to survive in such a reducing environment like C--rich
circumstellar envelopes of the 21$\mum$ sources
\citep{Posch2004ApJ...616}.
Therefore, Fe$_2$O$_3$ and Fe$_3$O$_4$
are unlikely responsible for the 21$\mum$ feature seen in PPNe.

\subsubsection{FeO}
\citet{Posch2004ApJ...616} pointed out that iron monoxide
(FeO; w\"ustite) can survive in the C-rich reducing environment
and proposed that nano-sized FeO dust may be the carrier
of the 21$\mum$ feature seen in PPNe.
FeO will be reduced to metal iron by UV photons in PNe,
while it is less likely for oxygen to stick to iron to form FeO in
AGB stars whose circumstellar dust is much hotter than in PPNe.
Since FeO can be either oxygenated to higher oxides of iron or
reduced to iron atoms, it can survive only in a very strict physical
and chemical environment, which is consistent with the observational
fact that the 21$\mum$ band is rarely seen and detected only in PPNe
that live for a short transitory period.

We adopt the dielectric functions of pure FeO measured by
\citet{Henning1997A&A...327} at temperatures $T$\,=\,10, 100, 200
and 300$\K$ to calculate the absorption spectrum of FeO dust. It is
found that FWHM of the 21$\mum$ band of FeO decreases from
$\simali$3.6$\mum$ at room temperature to $\simali$2.4$\mum$ at
$T$\,=\,100$\K$, while the band peak $\lambda_{\rm peak}$ shifts
from $\simali$19.9$\mum$ to $\simali$20.1$\mum$. At $T$\,=\,100$\K$,
both FWHM and $\lambda_{\rm peak}$ of FeO agree well with the
observed 21$\mum$ profile.

Adopting the dielectric functions of FeO measured at $T=100\K$ and
assuming spherical dust of radii $a= 1$\,nm as suggested by
\citet{Posch2004ApJ...616}, we calculate $\kabsint \approx
1.07\cm^3\g^{-1}$ for FeO (Fig.\,\ref{fig:kabs_21um}h shows the
calculated $\kabs$ profile which is closely reproduced by a Drude
function with $\lambdap$\,=\,20.1$\mum$ and $\gamma$\,=\,2.4$\mum$).
Nano-sized FeO grains will be stochastically heated by single
stellar photons (see \citealt{Draine_Li2001ApJ...551..807}). With a
Debye temperature of $\Theta \approx 650\K$
\citep{Radwanskia2008PhyB...403}, the peak temperature of a 1\,nm
FeO dust is $T_{\rm peak} \approx 143\K$ when heated by a typical
photon of 3\,eV in HD\,56126. At $\Td = 143\K$ the minimum abundance
requirement is $\Femin \approx 5.76\times10^{-7}$.
This is smaller than the stellar abundance
$\Festar \approx3.24\times10^{-6}$
(see Fig.\,\ref{fig:X2Hmin}). 
  Note that HD\,56126 is the most iron-poor
  object among the sixteen 21$\mum$ sources.
  Therefore, the FeO hypothesis will generally satisfy
  the abundance requirement.

FeO has no noticeable secondary features in the IR, except a small
shoulder in the blue wing of the 21$\mum$ band which is not seen in
the 21$\mum$ sources.

Stochastically-heated FeO nano dust is expected to have a
distribution of temperatures ($T < T_{\rm peak}=143\K$). Although
the optical properties of FeO change strongly with temperature, at
$T<100\K$ they are much less sensitive to temperature (see
\citealt{Henning1997A&A...327}). It is therefore reasonable to
expect that FeO nano dust with a distribution of temperatures at
$T<143\K$ also fits the observed 21$\mum$ profile since the
experimental spectrum of FeO obtained at $T=100\K$ provides the best
match.

Finally, one may ask how FeO is formed in the C-rich shells around
the 21$\mum$ sources where it is generally believed that all O atoms
are trapped in CO; or in other words, are there enough O atoms left
to form FeO? Our answer is ``yes''. The C and O abundances (relative
to H) of HD\,56126 are respectively $\Cstar\approx 447\ppm$ and
$\Ostar\approx 468\ppm$ \citep{VanWinckel2000A&A...354}. A
substantial fraction of the C atoms is required to be tied in the
carbonaceous dust components (amorphous carbon [a-C] and HAC) to
account for the bulk of the IR emission of HD\,56126 (see
\citealt{Hony...2003A&A...402..211H}). Taking $M_{\rm a-C} \approx
3.6\times 10^{-4}\,M_\odot$ and $M_{\rm HAC} \approx 3.6\times
10^{-4}\,M_\odot$ for the mass of amorphous carbon and HAC in the
dust shell around HD\,56126 \citep{Hony...2003A&A...402..211H}, we
estimate the amounts of C atoms locked in amorphous carbon and HAC
to be $\CtoH_{\rm a-C}\approx 68\ppm$
and $\CtoH_{\rm HAC}\approx 65\ppm$, respectively.\footnote{%
  Here we adopt $M_{\rm H}\approx 0.44\,M_\odot$,
  the high end of the H mass of the HD\,56126 shell.
  If we adopt the low end of $M_{\rm H}\approx 0.16\,M_\odot$,
  the required C depletion in dust needs to be increased by
  a factor of $\simali$2.8. This implies that there will be
  fewer C atoms in the gas phase to form CO and therefore
  more O atoms will be available to form FeO.
  We take the molecular weight of HAC to be
  $\mu_{\rm HAC} \approx 12.5\,\mu_{\rm H}$
  (with ${\rm H/\left(H+C\right)\approx 0.35}$,
   \citealt{Hony...2003A&A...402..211H}).
  }
Therefore, there will be at most $\CtoH \approx 314\ppm$
C atoms in the gas phase to react with O. Assuming all
gas-phase C atoms are tied with O atoms to form CO,
there will be $\OtoH\approx 154\ppm$ O atoms left for
other O-containing molecules and dust species.\footnote{%
     \citet{Duley1980ApJ...240..950D} showed that the oxidation
     of metallic iron is a very efficient process
     provided that enough O$_2$ molecules are present.
     FeO is a preferential product of low-temperature
     oxidation (e.g. see \citealt{Roberts1961TransSoc..57..99},
     \citealt{Fehlner1970Oxidation...2..59D},
     while Fe$_2$O$_3$ and Fe$_3$O$_4$
     are more likely formed at higher temperatures
     (e.g. see \citealt{Gail1998cpmg.conf..303G}).
     \citet{Posch2004ApJ...616} argued that
     the low-temperature oxidation of very
     small iron grains -- grains composed of $<10^3$ atoms
     with a size of $a \le 1$\,nm
     could result in pure FeO dust (instead of a large
     metallic iron core and a tiny FeO mantle).
     }
Even assuming all Fe atoms are locked up in FeO dust,
FeO only needs $\OtoH\approx 3.24\ppm$ O atoms.

\section{Summary}\label{sec:summary}
We have quantitatively examined
eight inorganic carrier candidates and one organic candidate
for the 21$\mum$ emission band detected in sixteen PPNe,
by confronting the abundance and accompanying
IR emission features required or predicted by each model
with that observed.
We take HD\,56126, the strongest 21$\mum$ feature source,
as a testing case.
The principal results of this paper are the following
(see Table\,\ref{tab:res} for a summary):
\begin{enumerate}
\item Among the nine carrier candidates,
      five (TiC nanoclusters, fullerenes coordinated with Ti atoms,
      SiS$_2$,  doped-SiC, and SiO$_2$-mantled SiC dust)
      are readily ruled out because they fall short of
      either titanium, sulphur, or silicon.
      Even under the most optimal condition,
      the minimum Ti, S or Si abundance required to
      account for the observed 21$\mum$ feature strength
      still substantially exceeds the available amount in
      the 21$\mum$ sources.
\item Three candidates
      (carbon-silicon mixtures, Fe$_3$O$_4$ and Fe$_2$O$_3$)
      are ruled out because they produce strong secondary features
      (in addition to the 21$\mum$ feature) which are not seen
      in the 21$\mum$ feature sources.
\item FeO nano dust, closely matching the observed 21$\mum$
      emission feature neither exceeds the Fe abundance
      budget nor produces undetected secondary features.
      There are also plenty of O atoms in the dust shell around
      HD\,56126 to form FeO, although it is a C-rich environment.
      Therefore, nano-sized FeO seems to be a viable candidate.

\end{enumerate}

By taking an alternative approach based on abundance constraints
complementary to that based on spectral profile matching, our
results are consistent with that of \citet{Posch2004ApJ...616}
except (1) we have also ruled out two newly-suggested candidates
(i.e. fullerenes coordinated with Ti atoms, carbon-silicon
mixtures); (2) SiC core-SiO$_2$ mantle dust is more firmly ruled out
as a valid candidate; and (3) we provide further support to the FeO
hypothesis of \citet{Posch2004ApJ...616}. Our approach does not rely
on detailed spectral profile fitting which could be largely affected
by dust size, shape and clustering effects.

We have not applied the above tests to the organic candidates
(i.e. urea or thiourea, PAHs, and HAC) for three reasons:
(1) the optical properties of urea or thiourea are not known;
(2) although ten of the sixteen 21 $\mum$ sources
    show PAH emission features at the so-called
    3.3, 6.2, 7.7, 8.2, and 11.3$\mum$
    ``unidentified IR (UIR)'' bands,
    PAHs are unlikely a viable candidate since their 21$\mum$
    band strength is much weaker than that of the ``UIR'' bands; and
(3) HAC is an ill-characterized material; its optical properties
    are sensitive to the H/C and sp$^2$/sp$^3$ ratios
    (see
    \citealt{Furton_Witt1999ApJ...526..752,Jones1990QJRAS..31..567});
    its 21$\mum$ band strength (especially relative to
    its vibrational bands at $\simali$6--8$\mum$)
    is not well determined.
To us, PAH clusters probably deserve a detailed investigation.

\section{ACKNOWLEDGMENTS}
We thank the anonymous referee for his/her very useful suggestions.
We thank A. K. Speck for helpful discussions. B.W.J. and K.Z. are
supported in part by China's grants 2007CB815406, NSFC 10473003 and
NCET-05-0144. A.L. is supported in part by Spitzer Theory Programs,
the Spitzer Cycle 3 GO program P30403 and NSF grant AST 07-07866.

\bibliographystyle{mn2e}
\bibliography{abundance}

\begin{thebibliography}{}

\bibitem[\protect\citeauthoryear{{Andersen}, {Posch} \& {Mutschke}}{{Andersen}
  et~al.}{2005}]{Andersen2005ESASP.577..447}
{Andersen} A.~C.,  {Posch} T.,    {Mutschke} H.,  2005, in {Wilson} A.,  ed.,
  ESA Special Publication Vol.~577, {Pitfalls in the identification of the 21
  micron feature}.
pp 447--448

\bibitem[\protect\citeauthoryear{{Bernatowicz}, {Cowsik}, {Gibbons} Patrick
  C.and~{Lodders}, {Fegley} Bruce, {Amari} \& {Lewis}}{{Bernatowicz}
  et~al.}{1996}]{Bernatowicz1996ApJ...472..760B}
{Bernatowicz} T.,  {Cowsik} R.,  {Gibbons} Patrick C.and~{Lodders} K.,
  {Fegley} Bruce J.,  {Amari} S.,    {Lewis} R.~S.,  1996, ApJ, 472, 760

\bibitem[\protect\citeauthoryear{{Buss} Jr., {Tielens}, {Cohen}, {Werner},
  {Bregman} \& {Witteborn}}{{Buss} et~al.}{1993}]{Buss1993ApJ...415}
{Buss} Jr. R.~H.,  {Tielens} A.~G.~G.~M.,  {Cohen} M.,  {Werner} M.~W.,
  {Bregman} J.~D.,    {Witteborn} F.~C.,  1993, ApJ, 415, 250

\bibitem[\protect\citeauthoryear{{Chigai}, {Yamamoto}, {Kaito} \&
  {Kimura}}{{Chigai} et~al.}{2003}]{Chigai...2003ApJ...587..771C}
{Chigai} T.,  {Yamamoto} T.,  {Kaito} C.,    {Kimura} Y.,  2003, ApJ, 587, 771

\bibitem[\protect\citeauthoryear{{Cl{\'e}ment}, {Mutschke}, {Klein} \&
  {Henning}}{{Cl{\'e}ment} et~al.}{2003}]{Clement2003ApJ...594}
{Cl{\'e}ment} D.,  {Mutschke} H.,  {Klein} R.,    {Henning} T.,  2003, ApJ,
  594, 642

\bibitem[\protect\citeauthoryear{{Cox}}{{Cox}}{1990}]{Cox1990A&A...236L}
{Cox} P.,  1990, A{\&}A, 236, L29

\bibitem[\protect\citeauthoryear{{Draine} \& {Li}}{{Draine} \&
  {Li}}{2001}]{Draine_Li2001ApJ...551..807}
{Draine} B.~T.,  {Li} A.,  2001, ApJ, 551, 807

\bibitem[\protect\citeauthoryear{{Duley}}{{Duley}}{1980}]{Duley1980ApJ...240..%
950D}
{Duley} W.~W.,  1980, ApJ, 240, 950

\bibitem[\protect\citeauthoryear{{Fehlner} \& {Mott}}{{Fehlner} \&
  {Mott}}{1970}]{Fehlner1970Oxidation...2..59D}
{Fehlner} F.~P.,  {Mott} N.~F.,  1970, {Oxidation of Metal}, 2, 59

\bibitem[\protect\citeauthoryear{{Furton}, {Laiho} \& {Witt}}{{Furton}
  et~al.}{1999}]{Furton_Witt1999ApJ...526..752}
{Furton} D.~G.,  {Laiho} J.~W.,    {Witt} A.~N.,  1999, ApJ, 526, 752

\bibitem[\protect\citeauthoryear{{Gail} \& {Sedlmayr}}{{Gail} \&
  {Sedlmayr}}{1998}]{Gail1998cpmg.conf..303G}
{Gail} H.-P.,  {Sedlmayr} E.,  1998, in Chemistry and Physics of Molecules and
  Grains in Space. Faraday Discussions No. 109 {Inorganic dust formation in
  astrophysical environments}.
p.~303

\bibitem[\protect\citeauthoryear{{Gilra}}{{Gilra}}{1972}]{Gilra1972PhDT}
{Gilra} D.~P.,  1972, PhD thesis, University of Wiconsin-Madison

\bibitem[\protect\citeauthoryear{{Goebel}}{{Goebel}}{1993}]{Goebel1993A&A...27%
8}
{Goebel} J.~H.,  1993, A{\&}A, 278, 226

\bibitem[\protect\citeauthoryear{{Goebel} \& {Moseley}}{{Goebel} \&
  {Moseley}}{1985}]{Goebel1985ApJ...290L..35G}
{Goebel} J.~H.,  {Moseley} S.~H.,  1985, ApJ, 290, L35

\bibitem[\protect\citeauthoryear{{Grevesse}}{{Grevesse}}{1989}]{Grevesse1989AI%
PC..183....9G}
{Grevesse} N.,  1989, in AIP Conference Proceedings Vol. 183: Cosmic abundances
  of matter {The abundances of matter in the sun}.
pp 9--16

\bibitem[\protect\citeauthoryear{{Henning} \& {Mutschke}}{{Henning} \&
  {Mutschke}}{1997}]{Henning1997A&A...327}
{Henning} T.,  {Mutschke} H.,  1997, A{\&}A, 327, 743

\bibitem[\protect\citeauthoryear{{Henning} \& {Mutschke}}{{Henning} \&
  {Mutschke}}{2001}]{Henning2001AcSpe..57..815H}
{Henning} T.,  {Mutschke} H.,  2001, Spectrochimica Acta, 57, 815

\bibitem[\protect\citeauthoryear{{Hony}, {Tielens}, {Waters} \& {de
  Koter}}{{Hony} et~al.}{2003}]{Hony...2003A&A...402..211H}
{Hony} S.,  {Tielens} A.~G.~G.~M.,  {Waters} L.~B.~F.~M.,    {de Koter} A.,
  2003, A{\&}A, 402, 211

\bibitem[\protect\citeauthoryear{{Hony}, {Waters} \& {Tielens}}{{Hony}
  et~al.}{2001}]{Hony2001A&A...378L}
{Hony} S.,  {Waters} L.~B.~F.~M.,    {Tielens} A.~G.~G.~M.,  2001, A{\&}A, 378,
  L41

\bibitem[\protect\citeauthoryear{{Hony}, {Waters} \& {Tielens}}{{Hony}
  et~al.}{2002}]{Hony2002A&A...390..533H}
{Hony} S.,  {Waters} L.~B.~F.~M.,    {Tielens} A.~G.~G.~M.,  2002, A{\&}A, 390,
  533

\bibitem[\protect\citeauthoryear{{Hrivnak}, {Volk} \& {Kwok}}{{Hrivnak}
  et~al.}{2000}]{Hrivnak2000ApJ...535..275H}
{Hrivnak} B.~J.,  {Volk} K.,    {Kwok} S.,  2000, ApJ, 535, 275

\bibitem[\protect\citeauthoryear{{Hrivnak}, {Volk} \& {Kwok}}{{Hrivnak}
  et~al.}{2009}]{Hrivnak2009arXiv0902.0077H}
{Hrivnak} B.~J.,  {Volk} K.,    {Kwok} S.,  2009, ArXiv e-prints
  (astro-ph/0902.0077)

\bibitem[\protect\citeauthoryear{{Huffman}}{{Huffman}}{1989}]{Huffman1989IAUS.%
.135..329H}
{Huffman} D.,  1989, in {Allamandola} L.~J.,  {Tielens} A.~G.~G.~M.,  eds,
  Interstellar Dust Vol.~135 of IAU Symposium, {Pitfalls in Calculating
  Scattering by Small Particles}.
p.~329

\bibitem[\protect\citeauthoryear{{Jiang}, {Zhang} \& {Li}}{{Jiang}
  et~al.}{2005}]{Jiang2005ApJ...630L}
{Jiang} B.~W.,  {Zhang} K.,    {Li} A.,  2005, ApJL, 630, L77

\bibitem[\protect\citeauthoryear{{Jiang}, {Zhang} \& {Li}}{{Jiang}
  et~al.}{2009}]{JZL2009EPS}
{Jiang} B.~W.,  {Zhang} K.,    {Li} A.,  2009, Earth, Planets, and Space, in
  press (astro-ph/0812.2015)

\bibitem[\protect\citeauthoryear{{Jones}, {Duley} \& {Williams}}{{Jones}
  et~al.}{1990}]{Jones1990QJRAS..31..567}
{Jones} A.~P.,  {Duley} W.~W.,    {Williams} D.~A.,  1990, QJRAS, 31, 567

\bibitem[\protect\citeauthoryear{{Justtanont}, {Barlow}, {Skinner}, {Roche},
  {Aitken} \& {Smith}}{{Justtanont} et~al.}{1996}]{Justtanont1996A&A...309}
{Justtanont} K.,  {Barlow} M.~J.,  {Skinner} C.~J.,  {Roche} P.~F.,  {Aitken}
  D.~K.,    {Smith} C.~H.,  1996, A{\&}A, 309, 612

\bibitem[\protect\citeauthoryear{{Kimura}, {Ishikawa}, {Kurumada}, {Tanigaki},
  {Suzuki} \& {Kaito}}{{Kimura} et~al.}{2005}]{Kimura2005crystal...275}
{Kimura} Y.,  {Ishikawa} M.,  {Kurumada} M.,  {Tanigaki} T.,  {Suzuki} H.,
  {Kaito} C.,  2005, Journal of Crystal Growth, 275, 977

\bibitem[\protect\citeauthoryear{{Kimura} \& {Kaito}}{{Kimura} \&
  {Kaito}}{2003a}]{Kimura2003crystal...255}
{Kimura} Y.,  {Kaito} C.,  2003a, Journal of Crystal Growth, 255, 282

\bibitem[\protect\citeauthoryear{{Kimura} \& {Kaito}}{{Kimura} \&
  {Kaito}}{2003b}]{Kimura2003MNRAS.343..385K}
{Kimura} Y.,  {Kaito} C.,  2003b, MNRAS, 343, 385

\bibitem[\protect\citeauthoryear{{Kimura}, {Nuth} III \& {Ferguson}}{{Kimura}
  et~al.}{2005}]{Kimura2005ApJ...632}
{Kimura} Y.,  {Nuth} III J.~A.,    {Ferguson} F.~T.,  2005, ApJL, 632, L159

\bibitem[\protect\citeauthoryear{{Koike}, {Hasegawa}, {Asada} \&
  {Hattori}}{{Koike} et~al.}{1981}]{Koike...1981Ap&SS..79...77K}
{Koike} C.,  {Hasegawa} H.,  {Asada} N.,    {Hattori} T.,  1981, ApSS., 79, 77

\bibitem[\protect\citeauthoryear{{Kraus}, {Nuth} III \& {Nelson}}{{Kraus}
  et~al.}{1997}]{Kraus1997A&A...328}
{Kraus} G.~F.,  {Nuth} III J.~A.,    {Nelson} R.~N.,  1997, A{\&}A, 328, 419

\bibitem[\protect\citeauthoryear{{Kwok}, {Volk} \& {Bernath}}{{Kwok}
  et~al.}{2001}]{Kwok2001ApJ...554L..87K}
{Kwok} S.,  {Volk} K.,    {Bernath} P.,  2001, ApJL, 554, L87

\bibitem[\protect\citeauthoryear{{Kwok}, {Volk} \& {Bidelman}}{{Kwok}
  et~al.}{1997}]{Kwok...1997ApJS..112..557K}
{Kwok} S.,  {Volk} K.,    {Bidelman} W.~P.,  1997, ApJS., 112, 557

\bibitem[\protect\citeauthoryear{{Kwok}, {Volk} \& {Hrivnak}}{{Kwok}
  et~al.}{1999}]{Kwok1999IAUS..191}
{Kwok} S.,  {Volk} K.,    {Hrivnak} B.~J.,  1999, in {Le Bertre} T.,  {Lebre}
  A.,   {Waelkens} C.,  eds, IAU Symp. 191: Asymptotic Giant Branch Stars {On
  the Origin of the 21 Micron Feature in Post-AGB Stars}.
p.~297

\bibitem[\protect\citeauthoryear{{Kwok}, {Volk} \& {Hrivnak}}{{Kwok}
  et~al.}{2002}]{Kwok2002ApJ...573}
{Kwok} S.,  {Volk} K.,    {Hrivnak} B.~J.,  2002, ApJ, 573, 720

\bibitem[\protect\citeauthoryear{{Kwok}, {Volk} \& {Hrivnak}}{{Kwok}
  et~al.}{1989}]{Kwok...1989ApJ...345L..51K}
{Kwok} S.,  {Volk} K.~M.,    {Hrivnak} B.~J.,  1989, ApJ, 345, L51

\bibitem[\protect\citeauthoryear{{Lattimer}, {Schramm} \&
  {Grossman}}{{Lattimer} et~al.}{1978}]{Lattimer1978ApJ...219..230L}
{Lattimer} J.~M.,  {Schramm} D.~N.,    {Grossman} L.,  1978, ApJ, 219, 230

\bibitem[\protect\citeauthoryear{{Li}}{{Li}}{2003}]{Li...2003ApJ...599L..45L}
{Li} A.,  2003, ApJL, 599, L45

\bibitem[\protect\citeauthoryear{{Li}}{{Li}}{2008}]{Li2008Optical}
{Li} A.,  2008, in {Mann} I.,  {Nakamura} A.,   {Mukai} T.,  eds, Small Bodies
  in Planetary Sciences (Lecture Notes in Physics vol. 758) {Optical Properties
  of Dust}.
pp 167--188

\bibitem[\protect\citeauthoryear{{Li}, {Shi} \& {Li}}{{Li}
  et~al.}{2008}]{Li2008MNRAS.391L..49L}
{Li} M.~P.,  {Shi} Q.~J.,    {Li} A.,  2008, MNRAS, 391, L49

\bibitem[\protect\citeauthoryear{{Nuth}, {Moseley}, {Silverberg}, {Goebel} \&
  {Moore}}{{Nuth} et~al.}{1985}]{Nuth1985ApJ...290L}
{Nuth} J.~A.,  {Moseley} S.~H.,  {Silverberg} R.~F.,  {Goebel} J.~H.,
  {Moore} W.~J.,  1985, ApJL, 290, L41

\bibitem[\protect\citeauthoryear{{Oudmaijer} \& {de Winter}}{{Oudmaijer} \& {de
  Winter}}{1995}]{Oudmaijer1995A&A...295L}
{Oudmaijer} R.~D.,  {de Winter} D.,  1995, A{\&}A, 295, L43

\bibitem[\protect\citeauthoryear{{Pierson}}{{Pierson}}{1996}]{Pierson1996}
{Pierson} H.~O.,  1996, {Handbook of Refractory Carbides and Nitrides:
  Properties, Characteristics, Processing and Applications}.
Noyes Publications, Westwood, NJ

\bibitem[\protect\citeauthoryear{{Posch}, {Mutschke} \& {Andersen}}{{Posch}
  et~al.}{2004}]{Posch2004ApJ...616}
{Posch} T.,  {Mutschke} H.,    {Andersen} A.,  2004, ApJ, 616, 1167

\bibitem[\protect\citeauthoryear{{Radwanski} \& {Ropka}}{{Radwanski} \&
  {Ropka}}{2008}]{Radwanskia2008PhyB...403}
{Radwanski} R.~J.,  {Ropka} Z.,  2008, Physica B Condensed Matter, 403, 1453

\bibitem[\protect\citeauthoryear{{Reddy}, {Lambert}, {Gonzalez} \&
  {Yong}}{{Reddy} et~al.}{2002}]{Reddy2002ApJ...564..482R}
{Reddy} B.~E.,  {Lambert} D.~L.,  {Gonzalez} G.,    {Yong} D.,  2002, ApJ, 564,
  482

\bibitem[\protect\citeauthoryear{{Roberts}}{{Roberts}}{1961}]{Roberts1961Trans%
Soc..57..99}
{Roberts} W.~M.,  1961, {Trans. Faraday Soc.}, 57, 99

\bibitem[\protect\citeauthoryear{{Rouleau} \& {Martin}}{{Rouleau} \&
  {Martin}}{1991}]{Rouleau1991JRASC..85..201R}
{Rouleau} F.,  {Martin} P.~G.,  1991, JRASC, 85, 201

\bibitem[\protect\citeauthoryear{{Sourisseau}, {Coddens} \&
  {Papoular}}{{Sourisseau} et~al.}{1992}]{Sourisseau1992A&A...254L}
{Sourisseau} C.,  {Coddens} G.,    {Papoular} R.,  1992, A{\&}A, 254, L1

\bibitem[\protect\citeauthoryear{{Speck} \& {Hofmeister}}{{Speck} \&
  {Hofmeister}}{2004}]{Speck2004ApJ...600}
{Speck} A.~K.,  {Hofmeister} A.~M.,  2004, ApJ, 600, 986

\bibitem[\protect\citeauthoryear{{Van Winckel} \& {Reyniers}}{{Van Winckel} \&
  {Reyniers}}{2000}]{VanWinckel2000A&A...354}
{Van Winckel} H.,  {Reyniers} M.,  2000, A{\&}A, 354, 135

\bibitem[\protect\citeauthoryear{{Volk}, {Kwok} \& {Hrivnak}}{{Volk}
  et~al.}{1999}]{volk...1999ApJ...516L..99V}
{Volk} K.,  {Kwok} S.,    {Hrivnak} B.~J.,  1999, ApJL, 516, L99

\bibitem[\protect\citeauthoryear{{Volk}, {Xiong} \& {Kwok}}{{Volk}
  et~al.}{2000}]{Volk2000ApJ...530}
{Volk} K.,  {Xiong} G.-Z.,    {Kwok} S.,  2000, ApJ, 530, 408

\bibitem[\protect\citeauthoryear{{Von Helden}, {Tielens}, {van Heijnsbergen},
  {Duncan}, {Hony}, {Waters} \& {Meijer}}{{Von Helden}
  et~al.}{2000}]{vonHelden...2000Sci...288..313V}
{Von Helden} G.,  {Tielens} A.~G.~G.~M.,  {van Heijnsbergen} D.,  {Duncan}
  M.~A.,  {Hony} S.,  {Waters} L.~B.~F.~M.,    {Meijer} G.,  2000, Science,
  288, 313

\bibitem[\protect\citeauthoryear{{Voshchinnikov} \& {Henning}}{{Voshchinnikov}
  \& {Henning}}{2008}]{Voshchinnikov2008A&A...483L...9V}
{Voshchinnikov} N.~V.,  {Henning} T.,  2008, A{\&}A, 483, L9

\bibitem[\protect\citeauthoryear{{Zhang}, {Jiang} \& {Li}}{{Zhang}
  et~al.}{2006}]{Zhang...2006PABei..24...43Z}
{Zhang} K.,  {Jiang} B.~W.,    {Li} A.~G.,  2006, Progress in Astronomy
  (Chinese) \, 24, 43

\bibitem[\protect\citeauthoryear{{Zhukovska} \& {Gail}}{{Zhukovska} \&
  {Gail}}{2008}]{Zhukovska2008A&A...486..229Z}
{Zhukovska} S.,  {Gail} H.-P.,  2008, A{\&}A, 486, 229

\end{thebibliography}

\clearpage

\clearpage

\begin{table}
\begin{center}
\caption{\label{tab:basicdata}
         Stellar and circumstellar parameters of HD\,56126}
\begin{tabular}{cccccc}
  \hline\hline
  \multicolumn{6}{l}{Central Star\footnotemark[1]} \\
  \hline
  $d$/kpc & $\Teff$/K & $R_\star/R_\odot$ & $L_\star/L_\odot$ \\
  2.4 & 7250 & 49.2 & 6054 & &\\ \hline
  \multicolumn{6}{l}{Dust Shell\footnotemark[1]} \\
  \hline
  $R_{\rm in}$/cm & $R_{\rm out}$/cm
                  & $M_{\rm H}$/$M_{\odot}$\footnotemark[2] &   \\
  4.5$\times$10$^{\rm 16}$ & 9.3$\times$10$^{\rm 16}$ &
  $\sim$0.16--0.44 & & & \\
  \hline
  \multicolumn{6}{l}{Abundance of Relevant Elements
                     (X/H, ppm)\footnotemark[3]} \\
  \hline
  Ti & S & Fe & C & O \\
  0.013 & 4.07 & 3.24 & 447 & 468 \\
  \hline
  \multicolumn{6}{l}{Total Emitted Power
                     in the 21$\mum$ Band
                     $\Etot$ ($\erg\s^{-1}$)\footnotemark[1]}\\
  \hline
  1.0$\times 10^{\rm 36}$ & & & & &\\ \hline
 \hline
\end{tabular}
\end{center}
\footnotemark[1] Data taken from \citet{Hony...2003A&A...402..211H}.
\footnotemark[2] Mass of the circumstellar shell depending on
                 the assumed gas-to-dust ratio.
\footnotemark[3] Data taken from \citet{VanWinckel2000A&A...354}.
\end{table}

\begin{table}
\begin{center}
\caption{\label{tab:res}
         Tests of abundance and possible accompanying
              features for nine carrier candidates
              of the 21$\mum$ feature seen in PPNe}

\begin{tabular}{ccccccc}
\hline
 Candidate& Element & $\kabsint$ & [X/H]$_{\rm min}$ &
[X/H]$_\star$ &
Associated & Pass ($\surd$) or \\
 Material& X &
($\cm^3\g^{-1}$)& (ppm) & (ppm)& Features& Fail ($\times$)\\

\hline
nano TiC  &  Ti    &  0.38 & 0.197 ($T=268\K$) & 0.013
          & ...    &  $\times$    \\
fullerenes\,+\,Ti  &  Ti  & 0.23 & 8.04 ($T=89\K$) & 0.013
          & ...    &   $\times$    \\
SiS$_2$   & S & 0.05 & 95.9 ($T=100\K$) & 4.07
          & 16.8$\mum$ &   $\times$    \\
doped-SiC & Si & 0.25 & 876 ($T=80\K$) & 8.50
          & 11.3$\mum$ & $\times$ \\
SiC\,+\,SiO$_2$ & Si & 0.21 & 805 ($T=70\K$)
                & 8.50 & 8.3,\,11.3$\mum$ & $\times$ \\
Si\,+\,C mixture & Si & ... & ... & ... & 9.5$\mum$ & $\times$ \\
Fe$_2$O$_3$ & Fe & 0.09 & 3.05 ($T=175\K$) & 3.24
            & 9.2,\,18,\,27.5$\mum$ & $\times$    \\
Fe$_3$O$_4$ & Fe & 0.06 & 2.92 ($T=200\K$) & 3.24
            & 16.5,\,24$\mum$ & $\times$    \\
FeO & Fe & 1.07 & 0.576 ($T=143\K$) & 3.24 & ... & $\surd$ \\

\hline
\end{tabular}
\end{center}
\end{table}

\end{document}